\documentclass[a4paper,11pt]{article}
\usepackage{pos}
\usepackage{subfig}
\usepackage{hyperref}
\hypersetup{urlcolor={blue},}

\usepackage{ulem}

\title{Probing singularities of Landau-gauge propagators with Pad\'e approximants}
\ShortTitle{Probing singularities of Landau-gauge propagators with Pad\'e approximants}

\author*[a,b]{Cristiane Yumi London}
\author[a]{Diogo Boito}
\author[a]{Attilio Cucchieri}
\author[a]{Tereza Mendes}

\affiliation[a]{Instituto de F\'isica de S\~ao Carlos, Universidade de S\~ao Paulo, \\
Caixa Postal 369, 13560-970 S\~ao Carlos, SP, Brazil}
\affiliation[b]{Grup de F\'isica Te\`orica, Departament de F\'isica,
Universitat Aut\`onoma de Barcelona, and Institut de F\'isica
d'Altes Energies (IFAE), The Barcelona Institute of Science and
Technology (BIST), Campus UAB, E-08193 Bellaterra (Barcelona), Spain}

\emailAdd{cristiane.london@usp.br}
\emailAdd{boito@ifsc.usp.br}
\emailAdd{attilio@ifsc.usp.br}
\emailAdd{mendes@ifsc.usp.br}

\abstract{Pad\'e approximants are employed in order to study the analytic structure of
the four-dimensional SU(2) Landau-gauge gluon and ghost propagators in the
infrared regime. The approximants, which are model independent, are used as
fitting functions to lattice data for the propagators, carefully propagating
uncertainties due to the fit procedure and taking into account all
possible correlations. Applying this procedure systematically to the
gluon-propagator data, we observe the presence of a pair of complex poles
at $p^2_{\mathrm{pole}} = (-0.37 \pm 0.05_{\mathrm{stat}}
\pm 0.08_{\mathrm{sys}}) \pm \, i\, (0.66 \pm 0.03_{\mathrm{stat}}
\pm 0.02_{\mathrm{sys}}) \, \mathrm{GeV}^2$, where ``stat'' represents
the statistical error and ``sys'' the systematic one. We also find a
zero on the negative real axis of $p^2$, at $p^2_{\mathrm{zero}} =
(-2.9 \pm 0.4_{\mathrm{stat}} \pm 0.9_{\mathrm{sys}}) \, \mathrm{GeV}^2$.
We thus note that our procedure --- which is based on a model-independent
approach and includes careful error propagation --- confirms the presence
of a pair of complex
poles in the gluon propagator, in agreement with previous works.
For the ghost propagator, the Padés indicate the existence of the single pole
at $p^2 = 0$, as expected.
We also find evidence of a branch cut on the negative real axis. Through
the use of the so-called D-Log Pad\'e method, which is designed to approximate
functions with cuts, we corroborate the existence of this cut for the ghost propagator.}

\FullConference{The 39th International Symposium on Lattice Field Theory,\\
8th-13th August, 2022,\\
Rheinische Friedrich-Wilhelms-Universität Bonn, Bonn, Germany}

\begin{document}
\maketitle

\section{Introduction}

Quantum chromodynamics (QCD), which is based on SU(3) Yang-Mills
theory, describes the strong interactions involving quarks
and gluons. One of its fundamental properties is color confinement, which
states that no color-charged particles can be found in isolation. Indeed,
up to now, no free quarks or gluons were observed in nature and hadronic
states are all colorless. Confinement in QCD is related to the fact that the
strong coupling, $\alpha_s$, runs with the energy, as predicted
by the QCD $\beta$-function; in particular, in the usual (perturbative)
$\overline{\rm{MS}}$ scheme, $\alpha_s$ assumes large values at low momenta,
i.e.\ in the infrared (IR) region, eventually
diverging at the Landau pole. A complete theoretical explanation
of color confinement is, however, still lacking.
As a first step in this direction, one can try to understand the behavior
of gluon and ghost propagators, which are the theory's fundamental degrees of
freedom, in the IR limit, i.e.\ going beyond the validity of
perturbative QCD.

Several theoretical frameworks were proposed to explain the
color confinement mechanism (in Landau gauge), such as the Gribov-Zwanziger
scenario~\cite{Gribov:1977wm,Vandersickel:2012tz} and the scaling solution of
Dyson-Schwinger equations~\cite{Alkofer:2000wg}, to name only two of them.
Both descriptions --- in their original
formulations --- predicted that the gluon propagator vanishes at zero momentum and
the ghost propagator has an enhanced singularity, i.e.\ it is more singular than
$1/p^2$ in the IR limit. These properties were verified with lattice
numerical simulations, using large lattice volumes, in 2 space-time
dimensions~\cite{Constraints,Modeling}.
On the contrary, in 3 and in 4 dimensions~\cite{Cucchieri:2007md,
Constraints,Bogolubsky:2009dc,Modeling}, lattice data showed that the ghost propagator, $G(p^2)$, is not
enhanced, i.e.\ $G(p^2 \approx 0) \sim 1/p^2$, and that the gluon
propagator does not go to zero at the origin.
These results for the propagators in the IR regime motivated novel
theoretical models that are in accordance with lattice simulations, such
as the decoupling solution of Dyson-Schwinger
equations~\cite{Aguilar:2008xm,Boucaud:2008ky}, the Refined Gribov-Zwanziger
scenario~\cite{Refined,Refined2,Dudal:2011gd} and the Curci-Ferrari
model~\cite{CF,Pelaez:2013cpa}.

Here, we report some of the results recently presented in
Ref.~\cite{Boito:2022rad}, to which we refer for further details.
The aim of our work is to study --- in a model-independent
way --- the analytic structure of the four-dimensional SU(2) (quenched)
Landau-gauge gluon and ghost propagators in the IR regime, by
using rational approximants as fitting functions to the data from
Refs.~\cite{Cucchieri:2007md,Cucchieri:2009xxr,Modeling}. We note that rational approximants were
recently applied by Falcão, Oliveira and
Silva~\cite{Falcao:2020vyr,Oliveira:2021job} to analyze the Landau-gauge
gluon and ghost propagators in the SU(3) case. We stress, however, that
in Refs.~\cite{Falcao:2020vyr,Oliveira:2021job} the issues related to
error propagation, and in particular the uncertainties arising from the fit
procedure, are not
fully discussed. In this work we propagated all the errors, considering
off-diagonal correlations when necessary. As shown in
Ref.~\cite{Boito:2022rad}, this limits considerably the number of
parameters that can be determined from our fits.

\section{Pad\'e Approximants}

The main type of rational approximants that we will employ as fitting functions
are the standard Padé approximants (PAs). We recall that a PA
$P_N^M(z)$ is a ratio of two polynomials of orders $M$ and
$N$~\cite{Baker1996pade,Baker1975essentials}, i.e.,
\begin{equation}
P_N^M(z) = \dfrac{Q_M(z)}{R_N(z)} =
\dfrac{a_0 + a_1 \, z + a_2 \, z^2 + \dots + a_M \, z^M}{1 + b_1 \, z + b_2 \, z^2 + \dots + b_N \, z^N} \; ,
\label{eq:pa}
\end{equation}
where we employed $b_0 = 1$ without loss of generality. The canonical method of
applying PAs is to approximate a function $f(z)$ whose Taylor series is known.
The Pad\'e coefficients are then obtained through a matching to the
Taylor-series coefficients of $f(z)$ up to order $N+M$, upon the expansion
of the Pad\'e. Thus, the PA will reproduce the first $N+M+1$ Taylor
coefficients of the function $f(z)$.

The most prominent advantage of employing PAs is that they allow for a
systematic and model-independent analysis. Moreover, which is
especially relevant to our work, they can reproduce analytic features of the
function they are approximating, such as poles, residues and zeros. In
addition, in some cases, the use of these approximants is validated by
convergence theorems. The most important one for this work is Pommerenke's
theorem, which states that PAs of the sequence $P_N^{N+k}(z)$, for fixed $k$,
applied to a meromorphic function $f(z)$, converges to $f(z)$ for
$N\to \infty$~\cite{Baker1996pade,Baker1975essentials}. On the other hand,
this theorem predicts that the Pad\'e can have extraneous poles, which
cannot be identified as singularities of the original function and
move away when the order of the PA is increased. There are also spurious
poles that can appear close to zeros, which are the \textit{defects} or
\textit{Froissart doublets}~\cite{Baker1996pade,Baker1975essentials,
MasjuanQueralt:2010hav} --- these defects are of transient nature and should
also disappear for PAs of higher order. It is important to
stress that approximants with such defects can still be employed
to study the considered function, away from these singularities.

In this work, we are going to use the Pad\'e approximants as fitting functions
to data, which departs from the standard Pad\'e theory and is not supported by
convergence theorems. However, experience shows that this approach
is rather powerful and that its application to similar
particle-physics problems, such as the extraction of resonance pole
positions, is very successful~\cite{Masjuan:2008fv,Masjuan:2012wy,
Masjuan:2013jha,Escribano:2013kba,Masjuan:2015cjl,Caprini:2016uxy,
VonDetten:2021rax}.
We also recall that the gluon and ghost propagators are expected to have
branch cuts in the complex plane and that, even though there are no
general theorems for Pad\'es applied to functions with cuts, it is
observed that the approximants mimic the branch cut of the function by
accumulating interleaved poles and zeros along the
cut~\cite{Baker1996pade,Baker1975essentials,MasjuanQueralt:2010hav,
Costin:2021bay}. A classic example in this context is the approximation of
the function $\log(z)$ by PAs.

\section{Results}

The lattice data fitted in this work have been previously presented
in Refs.~\cite{Cucchieri:2007md,Cucchieri:2009xxr,Modeling}
where more technical details can be found.
(See also Ref.~\cite{Boito:2022rad}.)
In particular, we use the data from a (symmetric) lattice with volume
$V = n^4 = 128^4$.
The lattice parameter was taken
to be $\beta = 2.2$, which leads to a lattice spacing $a$ of approximately
$0.210~\text{fm}$, considering $\sigma^{1/2} = 0.44~\text{GeV}$ for the string
tension~\cite{Bloch:2003sk}. Hence, the physical lattice volume is
about $(27~\text{fm})^4$, which can be essentially considered as
infinite volume, and the lowest non-zero physical momentum allowed is
$p_{min} = 2\,a^{-1} \sin(\pi / n ) \sim 46~\text{MeV}$.

The ghost-propagator data are considered in terms of the unimproved lattice
momenta $p^2 \, = \, \sum_{\mu} \, p_{\mu}^2$, where
$p_{\mu} = 2 \sin(\pi \hat{p}_{\mu}/n)$ and
$\hat{p}_{\mu} = 0, 1, \dots, n-1$. On the other hand, the gluon
propagator $D(p^2)$ is given in terms~\cite{Ma:1999kn} of the improved
momenta $p^2 \, = \, \sum_{\mu} \, p_{\mu}^2 \, + \, \frac{1}{12} \,
\sum_{\mu} \, p_{\mu}^4$, for the sake of reducing the effects due to
rotational-symmetry breaking, which are stronger at large
momenta. Thus, this definition mostly affects the values of momenta
outside the IR limit.

As shown in Ref.~\cite{Boito:2022rad}, the perturbative behavior of the
gluon propagator sets in around $2.0~\text{GeV}$; at the same time,
the ghost-propagator data are essentially perturbative for momenta
higher than $1.5~\text{GeV}$. Hence, in our work we focus the Pad\'e
analysis mostly in the IR region.

\subsection{Gluon propagator}

We start by employing the Pad\'e approximants as fitting functions to the
four-dimensional SU(2) Landau-gauge gluon propagator data. The fit parameters
are obtained through a $\chi^2$ minimization, taking into account
off-diagonal correlations when necessary. The fit uncertainties
are calculated using four different methods: Hessian matrix, Monte Carlo error
propagation, $\Delta \chi^2$ variation and linear error propagation.
We checked that results obtained with these four methods are
in good agreement. The errors presented in this work were calculated using
the Hessian matrix. As for the fit quality, it is judged by the
minimum $\chi^2$, divided by the degrees of freedom (dof), and by the
associated $p$-value. We stress that we limited the number of fit
parameters to 7, because for Pad\'es with higher order the statistical
uncertainty grows considerably and the fit is not meaningful anymore.
Also, the fits are performed in the region of
$\sqrt{p^2} < 2.4~\mathrm{GeV}$, and we have verified that the correlation
between the data points is negligible, so that it can be disconsidered in
the calculations.

\begin{figure}[!t]
\centering
\subfloat[]{\label{fig:pasdata}{\includegraphics[width=0.337\textwidth]{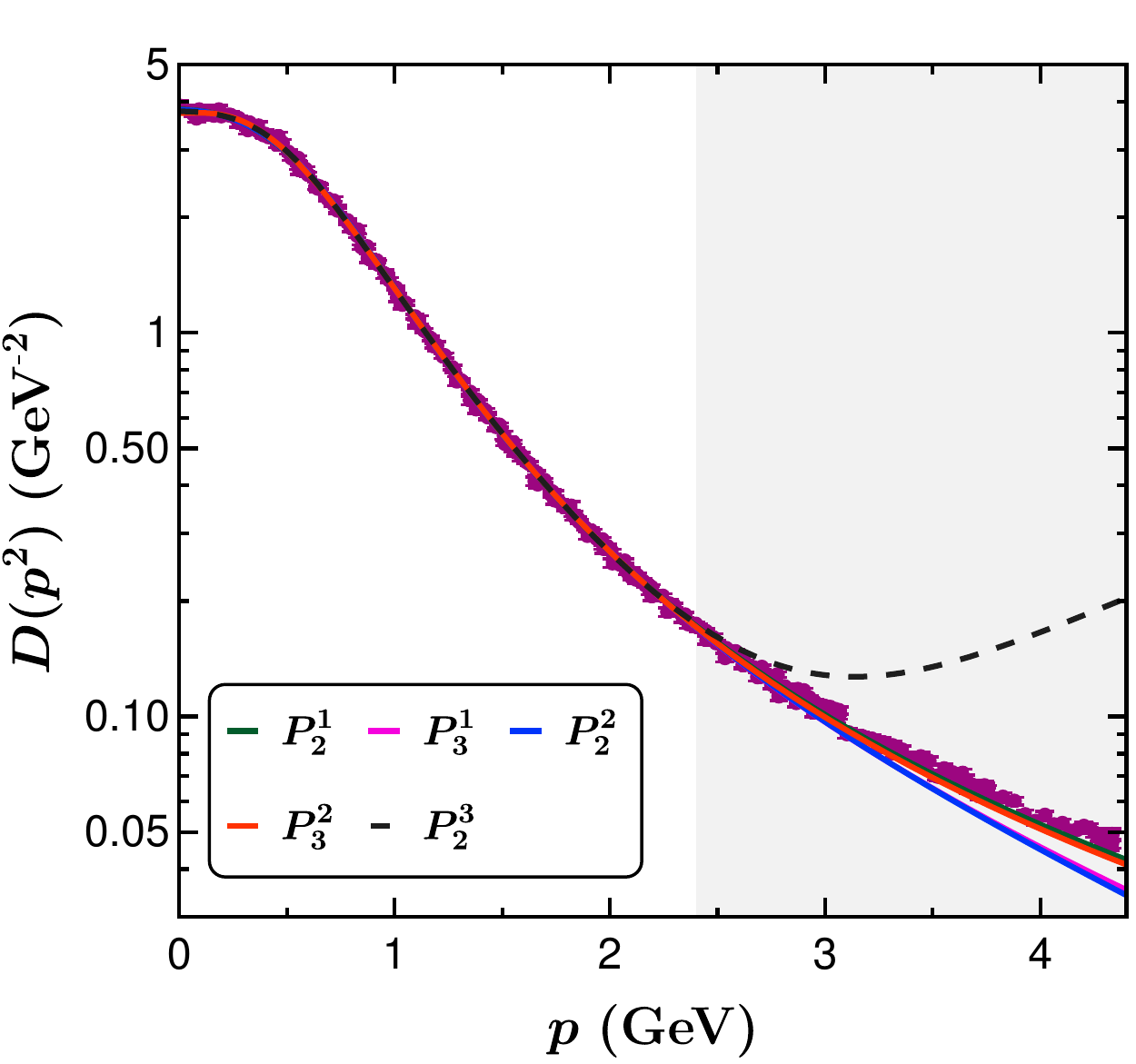}}}\hfill
\subfloat[]{\label{fig:paspoles}{\includegraphics[width=0.36\textwidth]{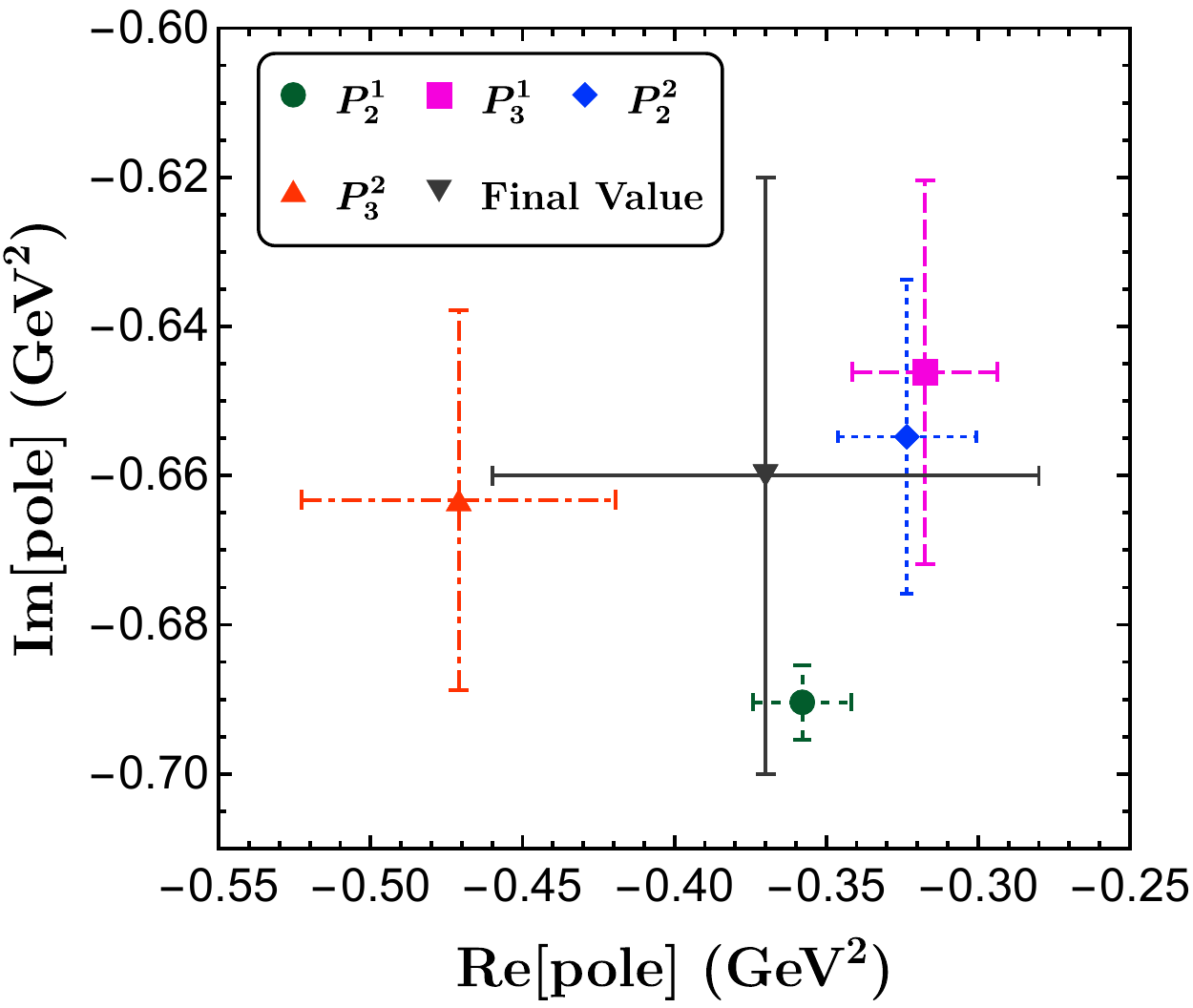}}}\hfill
\subfloat[]{\label{fig:paszeros}{\includegraphics[width=0.30\textwidth]{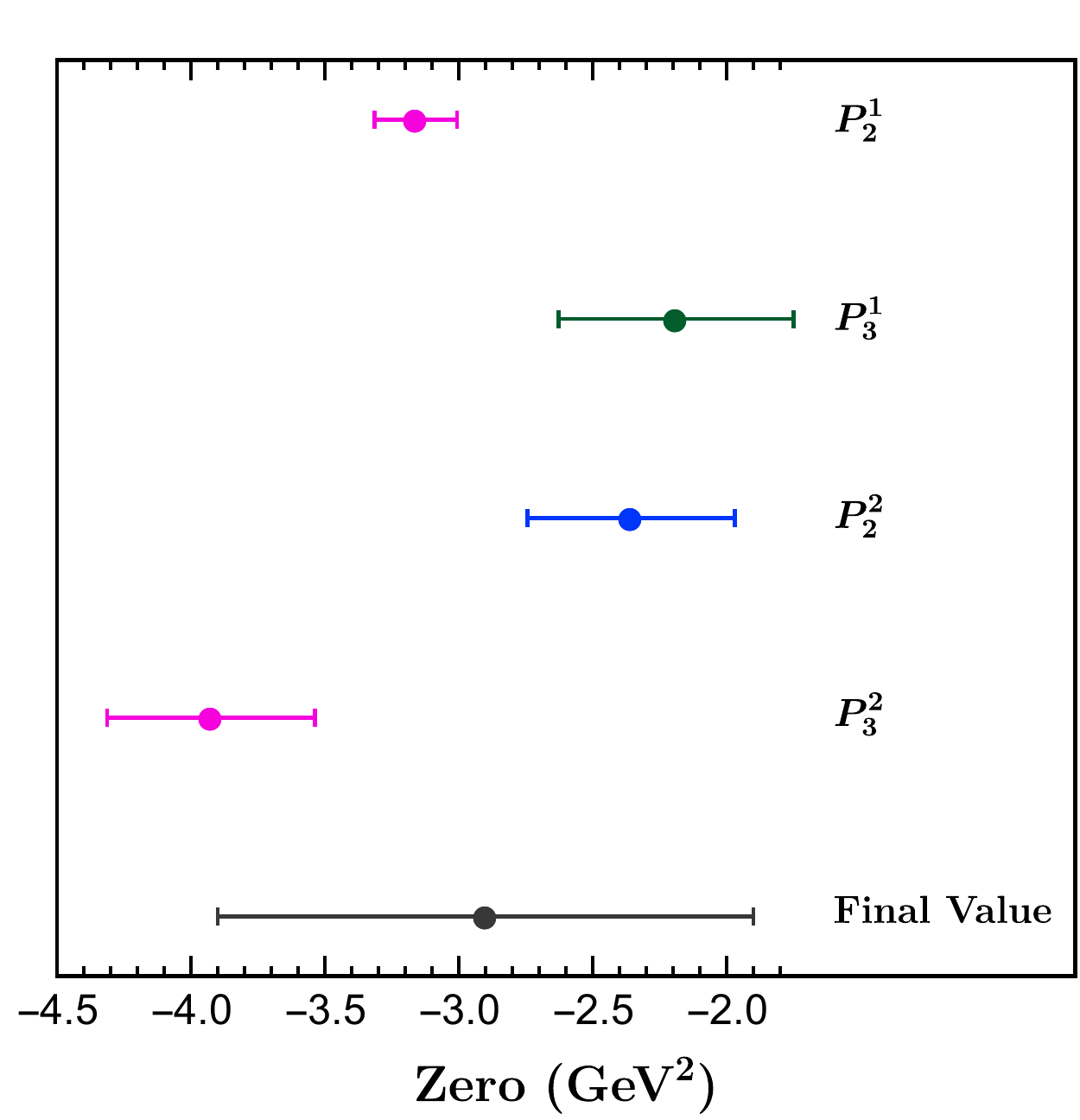}}}
\caption{Pad\'e approximants fitted to the Landau-gauge
gluon-propagator data and used to determine the final results
for poles and zeros. In (a) we show the comparison of the PAs
(used to determine the final results) and the lattice data;
the shaded region is not included in the fits.
In (b), respectively (c), we present the poles, respectively
the zeros, of each PA.
In both cases we show our final prediction in gray.}
\end{figure}

The following Pad\'e sequences were used to fit the lattice data:
$P_k^k(p^2)$, $P_k^{k+1}(p^2)$, $P_{k+1}^k(p^2)$ and
$P_{k+2}^k(p^2)$. The PAs that pass all reliability
tests,\footnote{See Ref.~\cite{Boito:2022rad} for more details.}
together with the lattice data, are presented in
Fig.~\ref{fig:pasdata}, where the white region is the fit window. Note
that, except for $P_2^3(p^2)$, all the approximants follow the expected
behavior of the propagator in the ultraviolet region.
The behavior at high energies of $P_2^3(p^2)$ can be explained by the fact
that this Pad\'e goes as $a_4 \, p^2$ for large $p^2$, with $a_4 > 0$ (albeit
small).
Due to the bad behavior of this PA, we did not use it in
our final estimates.

The pole position for each approximant is shown in Fig.~\ref{fig:paspoles}
and one can notice that the PAs have a consistent pair of complex poles. Our
final value is obtained as follows: the central value is the average of all
the Pad\'es results, the statistical uncertainty is the largest one from
the PAs, and the systematic error is half the maximum spread between
results from two different PAs. Employing this procedure we find that the final
position for the complex poles is
\begin{align}
\label{eq:polegluonfinal}
p_{\rm pole}^2 =[(-0.37 \pm 0.05_{\rm stat} &\pm 0.08_{\rm sys}) \,\, \pm \nonumber \\
& \pm \,\, i\,(0.66 \pm 0.03_{\rm stat} \pm 0.02_{\rm sys})] \,\, \mathrm{GeV}^2 \;,
\end{align}
where the first error is statistical and the second systematic. This final
value, which is indicated in gray in Fig.~\ref{fig:paspoles} (with the
errors added in quadrature), is in agreement with other results available
in the literature~\cite{Falcao:2020vyr,Oliveira:2021job,Modeling,
Dudal:2018cli}. Thus, the Pad\'es clearly favor a pair of complex
poles, with an imaginary part inconsistent with zero,
for the Landau-gauge gluon propagator.

Another noticeable feature in all the PAs of Fig.~\ref{fig:pasdata} is the
presence of a zero on the negative real axis of $p^2$. Applying the same method
used for the pole, we obtain the result
\begin{equation}
p_{\rm zero}^2 = (-2.9 \pm 0.4_{\rm stat} \pm 0.9_{\rm sys}) \,\, \mathrm{GeV}^2 \; ,
\label{eq:pazero}
\end{equation}
which is shown in Fig.~\ref{fig:paszeros} together with the zeros of the
considered approximants.

\subsection{Ghost propagator}

We now turn to applying the Pad\'e analysis to the Landau-gauge
ghost propagator. The method is mostly the same used for the gluon propagator
but, in this case, the correlation between the data points is
significant, reaching up to 0.75 in non-diagonal entries. It is known that fits
to strongly correlated data are problematic, since the covariance matrix
has small eigenvalues, which makes it hard to invert numerically.
What is more, the small eigenvalues generate huge numbers in the inverse
matrix. This engenders large fluctuations in the $\chi^2$ values, which are not
statistically meaningful, leading to biased~\cite{DAgostini:1993arp} or
unreliable~\cite{Boito:2011qt} results.
In order to avoid this problem, we will employ the so-called \textit{diagonal
fits}~\cite{Boito:2011qt}. In this procedure, only the diagonal covariance
matrix is used to determine the fit parameters; the corresponding fit
quality is denoted here as $Q^2$. All correlations are then included in the
error propagation according to the method explained in
Ref.~\cite{Boito:2011qt}. Let us stress that, since the diagonal
covariance matrix is used, the fit quality cannot be judged in absolute terms
by the value of $Q^2/\mathrm{dof}$.

By employing Pad\'e approximants to fit the lattice data of the ghost
propagator, it turns out that all the fit parameters of the
PAs are extremely large, of the order of $\mathcal{O}(10^{10})$ or
higher. This is caused by a pole very close to the origin, as can be
understood from Eq.~(\ref{eq:pa}).
We also verify that this feature persists when considering different
Pad\'e approximants, indicating that this pole is indeed
physical. For the sake of exploring additional analytical structures of the
ghost propagator, we then impose this pole at the origin, through
the so-called partial Pad\'e approximants
(PPAs)~\cite{Baker1996pade,Baker1975essentials},
which, in this case, are expressed as
\begin{equation}
\mathbb{P}^M_N(p^2) = \frac{Q_M(p^2)}{R_N(p^2) \, p^2} \; ,
\end{equation}
where $Q_M(p^2)$ and $R_N(p^2)$ are defined as before.

\begin{figure}[t]
\centering
\subfloat[]{\label{fig:pasdataghost}{\includegraphics[width=0.333\textwidth]{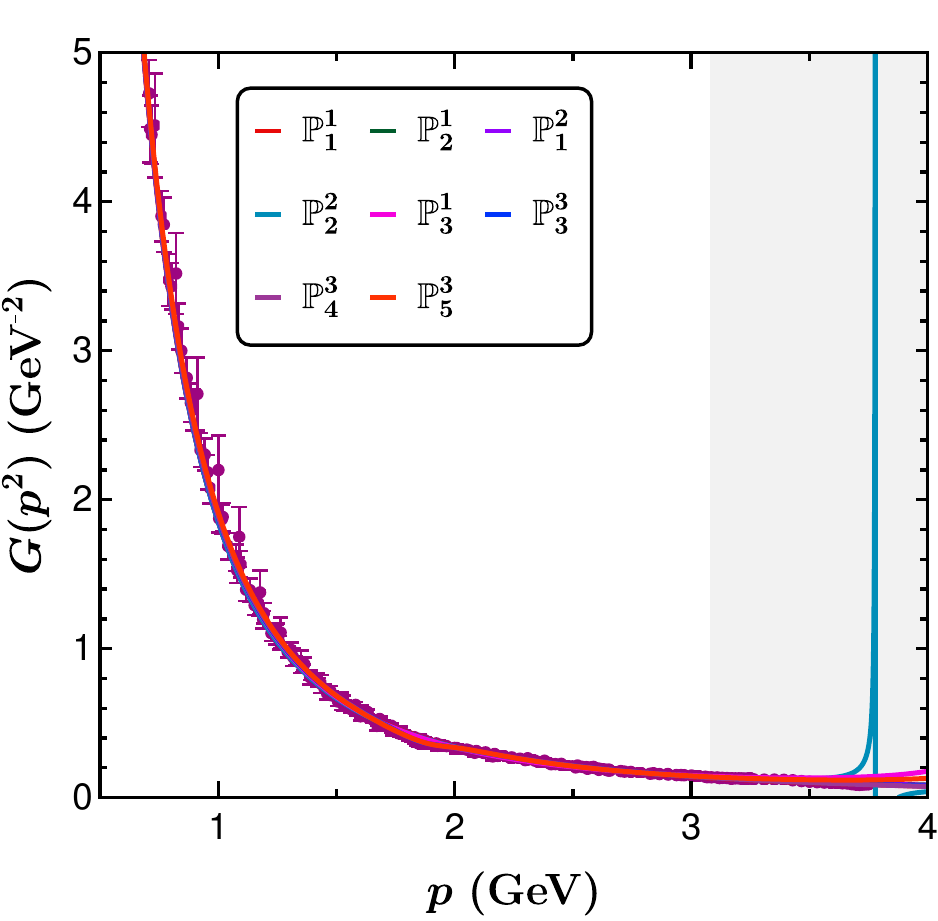}}}\hfill
\subfloat[]{\label{fig:paspolesghost}{\includegraphics[width=0.31\textwidth]{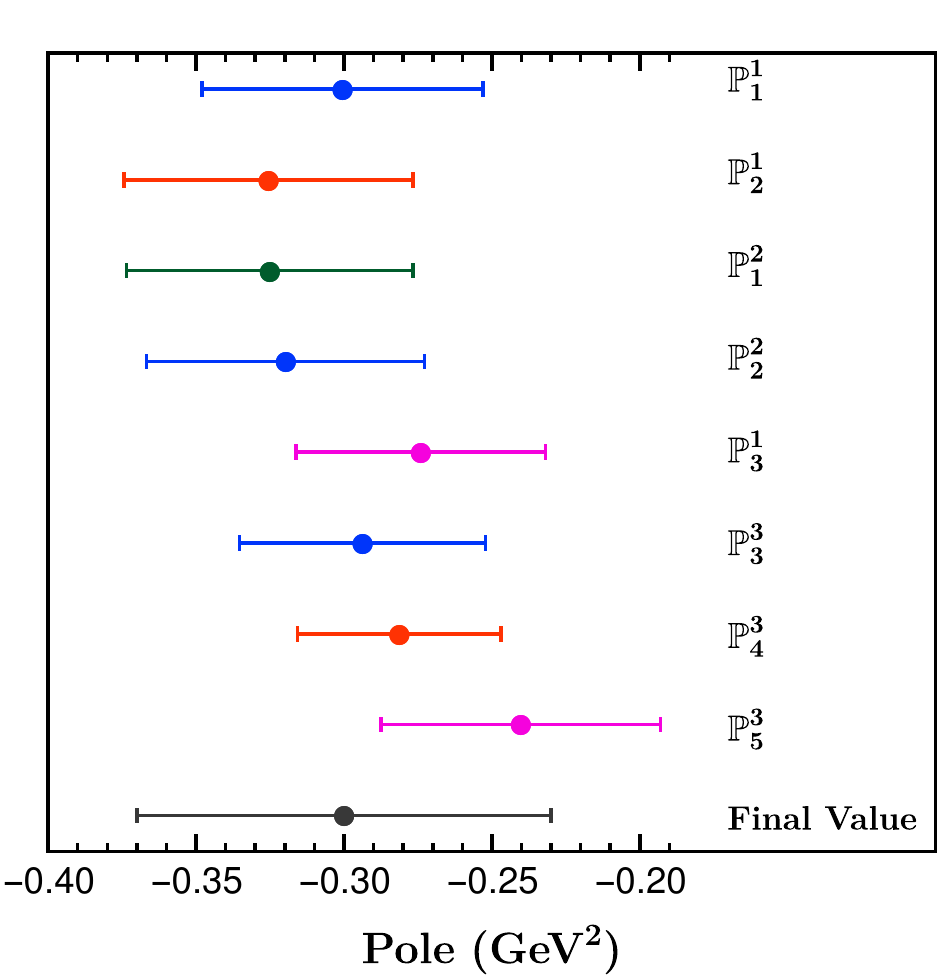}}}\hfill
\subfloat[]{\label{fig:paszerosghost}{\includegraphics[width=0.31\textwidth]{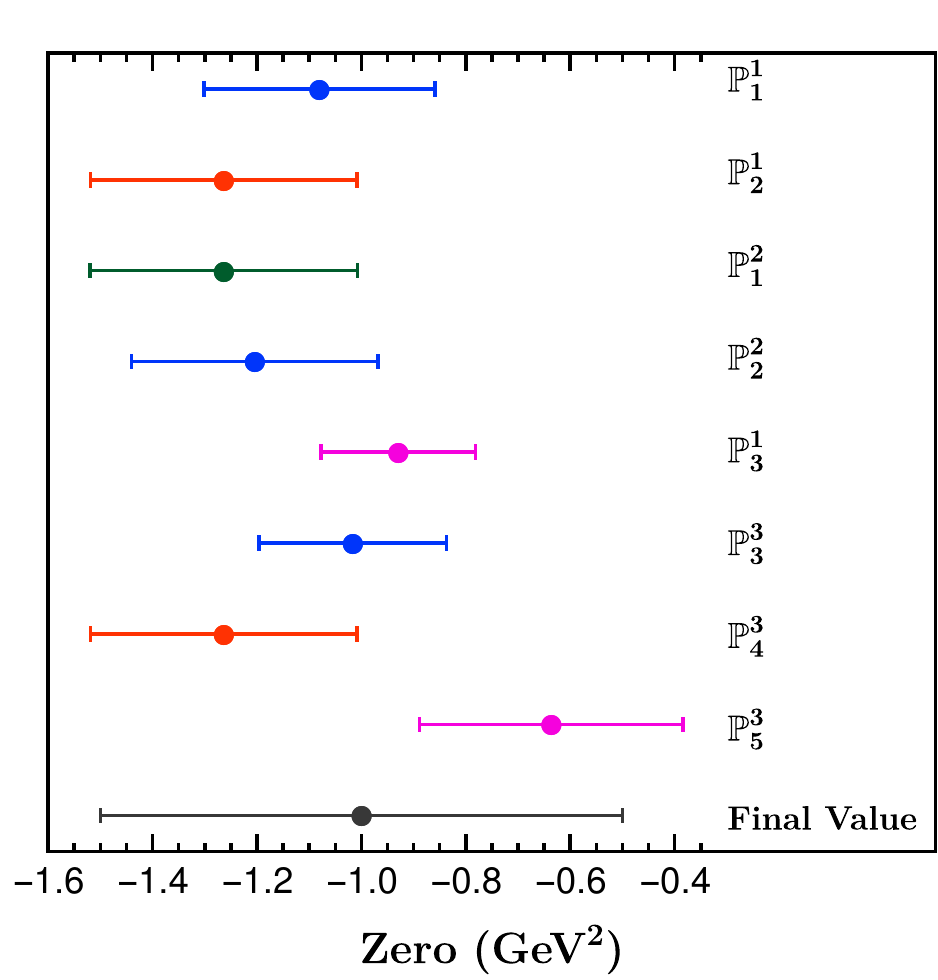}}}
\caption{Partial Pad\'e approximants fitted to the Landau-gauge
ghost-propagator lattice data used to determine our final results for poles
and zeros. In (a) we show the PPAs and the lattice data;
the shaded region is not included in the fits. In (b), respectively (c),
we present the poles, respectively the zeros, of each PPA.
In both cases, our final values are shown in gray at the bottom of each plot.
We recall that a pole (at about $p^2 = -0.30\,\mathrm{GeV}^2$)
followed by a zero (at about $p^2 = -1.0\,\mathrm{GeV}^2$) points towards
the presence of a branch cut along the negative real axis of $p^2$.}
\end{figure}

The fits were performed in the region $\sqrt{p^2} \leq 3.12~\mathrm{GeV}$
to avoid the appearance of Froissart doublets (in some PPAs) in the
considered range of momenta, which can spoil the extrapolation of the fit
results beyond the fit window. Indeed, as said above, these defects
may appear, but the approximants can still be used away from these
singularities.
We applied partial Pad\'es of the sequences $\mathbb{P}_k^k(p^2)$,
$\mathbb{P}_{k+1}^k(p^2)$, $\mathbb{P}_{k+2}^k(p^2)$ and
$\mathbb{P}_k^{k+1}(p^2)$ to fit the ghost-propagator data. As
before, the number of parameters of the PPAs that can be used to fit the data
is limited by the quality of the data. In particular, in this
case, partial Pad\'es up to order 8 can be used. The resulting PPAs and
the lattice data for the ghost propagator are shown in
Fig.~\ref{fig:pasdataghost}, where the gray region is not considered in the
fit. We note that $\mathbb{P}_2^2(p^2)$ also has a pole on the
positive real axis, which is almost cancelled by a near-by zero. This
singularity is a Froissart doublet and is a transient artifact of the PPA.

The Pad\'es also show a pole and a zero on the negative real axis of
$p^2$, which suggests a cut for the ghost propagator, considering that, as we
already stressed above, the Pad\'es emulate a cut by accumulating
interleaved poles and zeros.
Moreover, even though the uncertainties in the position of the poles and
zeros of PPAs of higher orders are large, it is possible to see that
their central values also present this pattern.
This corroborates the existence of a cut on the negative real axis.

The pole of each partial Pad\'e that passes all reliability tests are shown
in Fig.~\ref{fig:paspolesghost}. From these results we can determine our
final estimate, which is calculated by the same procedure employed in the
gluon-propagator analysis. This yields
\begin{equation}
\label{eq:pologhostfinal}
p^2_{\mathrm{pole}} = (-0.30 \pm 0.05_{\mathrm{stat}} \pm 0.05_{\mathrm{sys}}) \,\, \mathrm{GeV}^2.
\end{equation}
In addition, a zero is located at
\begin{equation}
p^2_{\mathrm{zero}} = (-1.0 \pm 0.3_{\mathrm{stat}} \pm 0.4_{\mathrm{sys}}) \,\, \mathrm{GeV}^2.
\end{equation}
The zeros of the PPAs together with our final value are shown in
Fig.~\ref{fig:paszerosghost}. As mentioned before, the appearance of interleaved
pole and zero can be an indication of a cut on the ghost propagator.

For the sake of better analyzing the existence of a cut on the
negative real axis, we also used the so-called D-Log Pad\'e
approximants~\cite{Baker1996pade,Baker1975essentials}. They are a suitable
alternative for functions with branch cuts because, instead of working with
the original function $f(z)$, which has a cut, one tries to
approximate a new function $F(z)$, which only has simple poles, and
then unfold the procedure. In particular, let us assume that the
function we are interested in is given
by~\cite{Baker1996pade,Baker1975essentials}
$f(z) = A(z) \, \frac{1}{(\mu - z)^\gamma} + B(z)$,
where $A(z)$ and $B(z)$ have a simple structure and are analytic at $z=\mu$,
and where $\gamma$ can be any real number. We then construct a new
function defined as~\cite{Baker1996pade,Baker1975essentials}
$F(z) = \frac{\mathrm{d}}{\mathrm{d} z}
\ln{f(z)} \approx \frac{\gamma}{(\mu - z)}$. Thus,
the branch point $\mu$ of $f(z)$ turns into a simple pole in $F(z)$,
whose residue is equal to $\gamma$. Hence, by unfolding the
procedure, the D-Log Pad\'e $\mathrm{Dlog}_N^M(z)$ of $f(z)$ is
given by~\cite{Baker1996pade,Baker1975essentials}
\begin{equation}
\mathrm{Dlog}_N^M(z) = f_{\mathrm{norm}}(0) \,
\exp{\left\{\int \mathrm{d}z' \,\, \bar{P}_N^M(z') \right\}} \; ,
\label{eq:dlogpa}
\end{equation}
where the Pad\'e $\bar{P}_N^M(z')$ is applied to the function $F(z)$ and
the constant $f_{\mathrm{norm}}(0)$ has to be adjusted in order to
reproduce the function at $z=0$, since the constant (zeroth-order)
term in the Taylor expansion of $f(z)$ is lost due to the derivative.

Clearly, in order to apply the D-log Pad\'es to the ghost propagator
$G(p^2)$, we need lattice data for the function $F(p^2)$, which is
the derivative of the logarithm of $G(p^2)$. This can be accomplished
by first taking the logarithm of the data and then calculating the
derivative, through finite differences. However, the standard
formulas for the derivative require equally spaced data points, a property
not satisfied by the ghost-propagator data.
Hence, to address this problem, the logarithm of the lattice data has been
linearly interpolated, to generate data points with a fixed separation of
$\Delta p^2 = 0.035~\mathrm{GeV}^2$; later, the derivative at a
given point is determined through the usual first-order formula. Of course,
this method introduces correlations between the data points, which
were thoroughly calculated and propagated. We note that, after
applying this procedure to the ghost-propagator data, the data points
have large uncertainties (in some cases larger than
100\%) with considerable statistical fluctuations.

\begin{figure}[t]
\centering
\subfloat[]{\label{fig:dlogdataghost}{\includegraphics[width=0.38\textwidth]{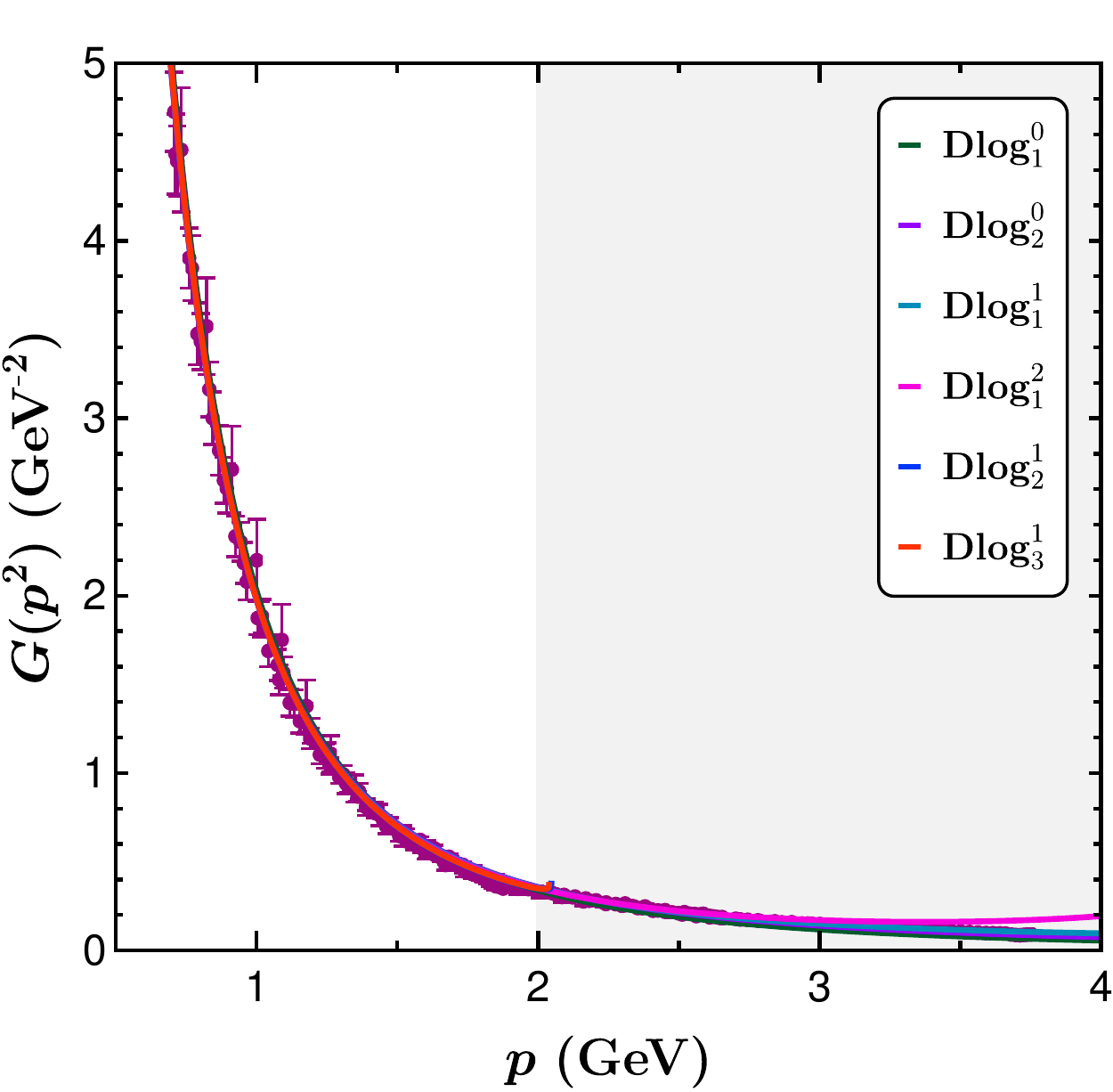}}}\hfill
\subfloat[]{\label{fig:poleconvghost}{\includegraphics[width=0.395\textwidth]{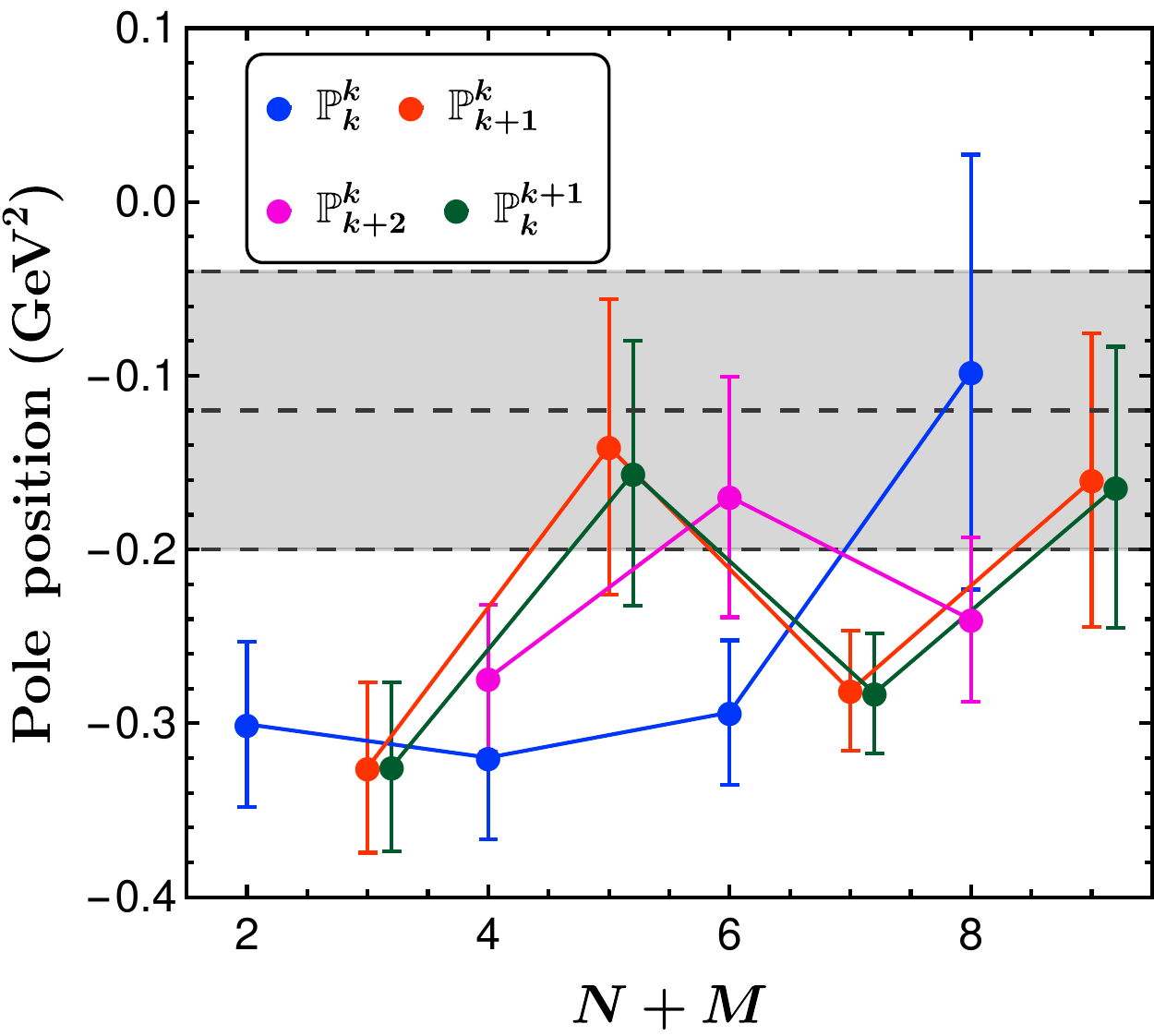}}}
\caption{(a) Lattice data for the ghost propagator and the D-Log
Pad\'es, built from the first-order numerical derivative of
the logarithm of the data. The shaded region is excluded from the fit.
(b) Comparison of the largest (real and nonzero) pole position --- from
four different partial Pad\'es sequences $\mathbb{P}^M_N(p^2)$ ---
with the branch-point position $p_c$ obtained from the D-Log
Pad\'es [see Eq.~(\ref{eq:pcfinal})] shown as the gray band.}
\end{figure}

For the D-Log Pad\'es, the region chosen for the fit is $\sqrt{p^2}
\leq 2~\mathrm{GeV}$, since the errors and fluctuations are
quite large for higher momenta; moreover, the data around
$\sqrt{p^2} = 2~\mathrm{GeV}$ are already in the perturbative region
(see Fig.~10a in Ref.~\cite{Boito:2022rad}).
We first apply the Pad\'es $\bar{P}_N^M(p^2)$ as fitting functions to the
prepared data, again employing the diagonal-fit method due to the
large correlation between the points.
Afterwards, Eq.~(\ref{eq:dlogpa}) is used to build the D-Log Pad\'es
belonging to the sequences: $\mathrm{Dlog}_k^k(p^2)$,
$\mathrm{Dlog}_{k+1}^k(p^2)$,
$\mathrm{Dlog}_{k+2}^k(p^2)$ and $\mathrm{Dlog}_k^{k+1}(p^2)$.
Due to the large uncertainties and fluctuations of the data points,
the maximum number of parameters was set to five.
The D-log-Pad\'e results are reported in
Fig.~\ref{fig:dlogdataghost}, together with the lattice data
(in purple). The fits have been performed considering only
the white region.
It is possible to see a good agreement with the data, even outside
the fit window. Moreover, every approximant of
Fig.~\ref{fig:dlogdataghost} has a singularity of the type
$(p_c-p^2)^{-\gamma}$,
where the branch point $p_c$ is always located on the negative real axis,
not too far from the pole determined by the partial Pad\'es [see
Eq.~(\ref{eq:pologhostfinal})], and the multiplicity $\gamma$ is not
compatible with one, which indicates that $p_c$ is not a simple pole.
Using these results, our final estimate for the branch-point
position is
\begin{equation}
\label{eq:pcfinal}
p_c^2 = (-0.12\pm 0.08_{\rm stat}\pm 0.02_{\rm sys})~\mathrm{GeV}^2 \; .
\end{equation}
It is important to emphasize that our systematic uncertainty for $p_c$ may
be underestimated, due to the fact that the evaluation of the numerical
derivative introduces an additional source of error. This result, together
with the exponent $\gamma$ not being compatible with one, reinforces
that the pole and zero of the PPAs are a manifestation of a cut on the negative
real axis.

Finally, let us compare the branch point from the D-log Pad\'es
with the pole prediction of the partial Pad\'es. We recall that,
if the cut at negative $p^2$ does exist, the largest (negative)
pole of the PPAs should indicate the position of the branch point.
In Fig.~\ref{fig:poleconvghost} the gray band corresponds to the
(above) branch point predicted by the D-log PAs; we
also show the largest non-zero pole position for different PPAs.
As one can notice, the central values of the pole are in
better agreement with $p_c$ for higher-order PPAs.
Also, the final results obtained for the branch point,
Eq.~(\ref{eq:pcfinal}), and for the pole extracted from the PPAs,
Eq.~(\ref{eq:pologhostfinal}), are compatible within $1.7~\sigma$.
Thus, these results are in good agreement and corroborate the existence
of a cut on the negative real axis for the ghost propagator.

\section{Conclusions}
\label{sec:conclusions}

In this work we applied a systematic and model-independent method to study the
analytic structure of the IR Landau-gauge gluon and ghost propagators,
by using rational approximants. In particular, Pad\'e approximants, partial
Pad\'e approximants and also D-log Pad\'e approximants were employed as fitting
functions to four-dimensional SU(2) lattice data~\cite{Cucchieri:2007md,
Cucchieri:2009xxr,Modeling} of both propagators.
We stress that Refs.~\cite{Falcao:2020vyr,Oliveira:2021job} also applied
Pad\'e approximants to fit SU(3) lattice propagators and our main conclusions
are in agreement with their results. Notice, however, that in
our work the errors were carefully propagated, considering all the
correlations. In addition, we also estimated the systematic
uncertainty of the employed method and explored other types of
approximants: the partial Pad\'es and the D-Log Pad\'es.

For the gluon propagator, the Pad\'e approximants presented evidence of a pair
of complex poles located at $p^2_{\mathrm{pole}} = [(-0.37 \pm 0.09) \pm
i \,(0.66 \pm 0.04)] \,\, \mathrm{GeV}^2$, where the errors are added in
quadrature. The PAs also indicated a zero on the negative real axis at
$p^2_{\rm zero}=(-2.9 \pm 1.0)~\text{GeV}^2$.

In the ghost-propagator case, the Pad\'es clearly show the existence of
a pole at the origin. We then imposed this information by using
the so-called partial Pad\'es to fit the ghost-propagator data.
These fits indicated the presence of a pole,
followed by a zero, on the negative real axis.
Higher-order PPAs also had additional interleaved poles and zeros along the
negative real axis, further suggesting the existence of a cut.
In order to investigate this result, we employed D-Log Pad\'es
and found a cut with branch point at $p_c^2 =(-0.12 \pm 0.08)~\text{GeV}^2$.
This value is compatible with the pole obtained using partial
Pad\'es, which is another corroboration for the existence of a
cut on the negative real axis for the ghost propagator.

\section{Acknowledgements}

DB's work was supported by the S\~ao Paulo Research Foundation (FAPESP) grant
No.\ 2021/06756-6 and CNPq grant No.\ 308979/2021-4. AC and TM acknowledge
partial support from FAPESP and CNPq. The work of CYL was financed by FAPESP
grants No.\ 2020/15532-1 and No.\ 2022/02328-2 and CNPq grant No.\ 140497/2021-8.

\end{document}